 \documentclass[final,5p,times]{elsarticle}

\usepackage{amssymb}
\usepackage{amsmath}
\usepackage{amsmath,amssymb,amsfonts}
\usepackage{algorithmic}
\usepackage{graphicx}
\usepackage{textcomp}
\usepackage{xcolor}

\usepackage{subcaption}
\usepackage{amsthm}
\usepackage{tabularx}
\usepackage{booktabs}
\usepackage{multirow}

\journal{}
\begin{document}
\begin{sloppypar}
\begin{frontmatter}



\title{BFLN: A Blockchain-based Federated Learning Model for Non-IID Data} 


\author[1,2]{Yang Li} 
	\author[2,3]{Chunhe Xia} 
	\author[1,2]{Dongchi Huang}
	\author[4]{Xiaojian Li} 
	\author[5]{Tianbo Wang} 

\affiliation[1]{
	organization={School of Computer Science and Engineering, Beihang University},
	city={Beijing},
	postcode={100191}, 
	country={China}
}
	\affiliation[2]{
	organization={Key Laboratory of Beijing Network Technology, Beihang University},
	city={Beijing},
	postcode={100191}, 
	country={China}
	  }
	\affiliation[3]{
	organization={Guangxi 	Collaborative Innovation Center of Multi-Source Information Integration and Intelligent Processing, Guangxi Normal University},
	city={Guilin },
	postcode={541004}, 
	country={China}
	  }
\affiliation[4]{
	organization={College of Computer Science and Information
		Technology, Guangxi Normal University},
	city={Guilin},
	postcode={541001}, 
	country={China}
}	
    \affiliation[5]{
    organization={	School of Cyber Science and Technology, Beihang University  },
    city={Beijing},
    postcode={100191}, 
    country={China}
  }


\begin{abstract}
As the application of federated learning becomes increasingly widespread, the issue of imbalanced training data distribution has emerged as a significant challenge. Federated learning utilizes local data stored on different training clients for model training, rather than centralizing data on a server, thereby greatly enhancing the privacy and security of training data. However, the distribution of training data across different clients may be imbalanced, with different categories of data potentially residing on different clients. This presents a challenge to traditional federated learning, which assumes data distribution is independent and identically distributed (IID). This paper proposes a Blockchain-based Federated Learning Model for Non-IID Data (BFLN), which combines federated learning with blockchain technology. By introducing a new aggregation method and incentive algorithm, BFLN enhances the model performance of federated learning on non-IID data. Experiments on public datasets demonstrate that, compared to other state-of-the-art models, BFLN improves training accuracy and provides a sustainable incentive mechanism for personalized federated learning.
\end{abstract}

%

\end{frontmatter}


\section{Introduction}
\label{Introduction}

Federated learning, introduced by Google in 2017 \cite{mcmahan2017communication}, aims to address the privacy and security concerns associated with traditional machine learning training data. It is now widely applied in fields such as transportation, healthcare, and finance, effectively protecting users' personal data privacy and security. In federated learning, different training clients typically have differently distributed training data. Traditional federated learning involves training the same model across all clients and using aggregation algorithms to produce a global model, which is updated through multiple rounds of parameter exchanges. This method of aggregating a single global model during training is also known as centralized federated learning \cite{li2020review}.
%
However, traditional centralized federated learning is a double-edged sword \cite{mammen2021federated, zhang2021survey, kairouz2021advances}. On the positive side, centralized federated learning consolidates all local models. In the initial federated learning aggregation algorithm, FedAvg, the parameters of local models are averaged, yielding good results on independently and identically distributed (IID) data. This aggregation method is relatively simple and achieves effective results. On the negative side, centralized federated learning only aggregates a single global model, which may not perform well on clients with differently distributed training data. Using the same global model for clients with different data distributions can lead to suboptimal performance, as these clients lack a targeted global model. When there are insufficient training samples, clients cannot learn effective data features through the aggregation of the global model, leading to the non-IID problem in federated learning.
For federated learning research, keeping training data local and sharing local model parameters not only protects the privacy and security of local training data but also provides clients with different data distributions a new way to learn more features from other clients' local data \cite{zhao2018federated}. This is particularly important in cases where sample distribution is imbalanced and individual clients have a small amount of training data.

To address the non-IID (non-independent and identically distributed) problem in federated learning, a solution known as personalized federated learning has been proposed \cite{tan2022towards}. The non-IID problem refers to the issue where centralized federated learning fails to meet the model training needs of each training client due to inconsistent local training data distributions across different clients \cite{zhu2021federated}. personalized federated learning offers targeted model updates for each training client, allowing them to train models based on their own data characteristics and meet their specific model update requirements.
However, as an emerging technology, personalized federated learning currently lacks effective incentive mechanisms to encourage active participation from training clients in the training process and achieve effective positive updates to the global model \cite{yan2022survey}. Without effective incentives, training clients may lack motivation for model training, and the necessary computational and communication resources for training may not be adequately compensated. Consequently, training clients may participate passively or not at all, hindering the continuous development of federated learning.
Compared to centralized federated learning, personalized federated learning does not rely on a single global model, making it challenging to uniformly evaluate the contributions of local training clients to the global model's training. Therefore, more equitable and scientific incentive mechanisms are needed to promote the sustained development of personalized federated learning.


To address the aforementioned issues, we propose a Blockchain-based Federated Learning Model for Non-IID Data (BFLN), which consists of two key components: the prototype-based aggregation algorithm (PAA) and the clustering centroids-based consensus algorithm (CCCA).

In PAA, we cluster local models using a prototype-based approach, where models within the same cluster are aggregated to achieve personalized updates. This effectively addresses the non-IID problem in federated learning. In CCCA, we designate the training client closest to the cluster centroid as the representative client for that cluster, which then joins a packing queue and takes turns packaging blocks, ensuring the sustainability of the blockchain packaging process. Additionally, CCCA introduces an incentive mechanism based on the number of cluster members: the more members in a cluster, the greater the rewards for its training clients. This is based on the premise that a larger number of members within a cluster indicates a greater contribution of that cluster's local models to the training of other local models, and thus, they deserve more rewards.

Experimental results demonstrate that our model not only effectively addresses the non-IID problem but also provides a novel solution for the incentive mechanism in personalized federated learning.

Our main contributions can be summarized as follows:
\begin{itemize}
	\item We proposed the BFLN model to address the non-IID problem in federated learning and the lack of effective incentive mechanisms in personalized federated learning. We designed a prototype-based aggregation algorithm and a clustering centroids-based consensus algorithm to solve these issues.
	\item We introduced a new personalized federated learning aggregation method by combining the Pearson correlation coefficient and spectral clustering techniques. This method outperforms the vanilla aggregation algorithm in handling the non-IID problem.
	\item We developed a consensus algorithm to ensure the sustainability of the blockchain packaging process. Within this consensus algorithm, we proposed an incentive mechanism tailored for personalized federated learning. By integrating federated learning with blockchain technology, we achieved a trustworthy distribution of tokens in the incentive mechanism.
	\item Experiments on three real-world datasets show that our model outperforms baseline works in non-iid data distribution's accuracy.
\end{itemize}

The rest of this article is organized as follows. In Section \ref{RELATED WORK}, we provide a comprehensive review of related work. Then, in Section \ref{PRELIMINARY}, we introduce the background knowledge. Section \ref{METHODS} details our proposed BFLN, including PAA and CCCA. In Section \ref{Experiments}, we conduct a series of experiments on public datasets to evaluate BFLN. Finally, Section \ref{Conclusion and Future Work}  summarizes this work and discusses future research directions.

\section{RELATED WORK}
\label{RELATED WORK}
This section provides a comprehensive review of relevant literature.  We start by briefly introducing the work on non-iid data distribution in personalized federated learning. Subsequently, we delve into the state-of-the-art approaches in incorporating incentive mechanisms into FL.

Many personalized federated learning (FL) approaches have been proposed to address the non-i.i.d. (non-independent and identically distributed) problem.
Canh T. Dinh et al. \cite{Dinh2020PersonalizedFL} proposed the personalized FL (pFedMe) algorithm using Moreau envelopes as clients' regularized loss functions. This algorithm helps decouple the optimization of personalized models from the learning of the global model in personalized FL's bi-level problem.
Alireza Fallah et al. \cite{Fallah2020PersonalizedFL} explored personalized solutions for federated learning to find an initial shared model that users can adapt to their local datasets by performing one or a few gradient descent steps on their local data.
A. Tan et al. \cite{Tan2021TowardsPF} summarized and analyzed the development of personalized federated learning, examining its motivations, highlighting its ideas, challenges, and opportunities, and offering prospects for its future development.
Liam Collins et al. \cite{Collins2021ExploitingSR} proposed a new federated learning framework to learn shared data representations across clients and local models for each client. This method demonstrated linear convergence in a linear setting, achieving near-optimal sample complexity and effectively reducing the dimensionality of each client's problem.
Yuyang Deng et al. \cite{Deng2020AdaptivePF} introduced the adaptive personalized federated learning (APFL) algorithm, where each client trains their local model while contributing to the global model. They also proposed a communication-efficient dual-stage optimization method to reduce the number of communication rounds.
Aviv Shamsian et al. \cite{Shamsian2021PersonalizedFL} proposed a new approach using a hypernetwork to tackle the non-i.i.d. problem by training a central hypernetwork model that generates a set of models, one for each client. This architecture provides effective parameter sharing among clients while maintaining the ability to generate unique and diverse personalized models. Additionally, since hypernetwork parameters are never transmitted, this method decouples communication costs from model size.
Qiong Wu et al. \cite{Wu2020PersonalizedFL} proposed a personalized federated learning framework for smart IoT applications in a cloud-edge architecture. To address heterogeneity in IoT environments, this study examined personalized federated learning methods to mitigate the negative impacts of heterogeneity and meet the requirements of rapid processing and low latency in smart IoT applications.
Michael Zhang et al. \cite{Zhang2020PersonalizedFL} calculated the optimal weighted model combination for each client based on how much a client can benefit from another client's model.
Xiaokang Zhou et al. \cite{Zhou2024PersonalizedFL} proposed a personalized federated learning framework based on model contrastive learning (PFL-MCL). This framework effectively enhances communication and interaction in the metaverse environment by leveraging large-scale, heterogeneous, and multi-modal metaverse data. Unlike traditional FL architectures, this framework develops a multi-center aggregation structure that learns multiple global models based on dynamically updated local model weight changes, and it constructs a hierarchical neural network structure that includes personalized and federated modules.
Zihan Chen et al. \cite{Chen2023PersonalizedFL} introduced a new PFL algorithm, FedACS, with an attention-based client selection mechanism. FedACS integrates an attention mechanism to enhance collaboration among clients with similar data distributions and alleviate data scarcity issues. It prioritizes and allocates resources based on data similarity.

Existing research has proposed numerous efficient personalized federated learning methods. However, there is still a lack of solutions that perform well under highly skewed training data distributions.

At the same time, various incentive mechanisms for federated learning have been proposed.
Peng Sun et al. \cite{Sun2021PainFLPP} introduced a contract-based personalized privacy-preserving incentive mechanism called Pain-FL. This mechanism provides customized payments to employees with different privacy preferences as compensation for privacy leakage costs while ensuring satisfactory convergence performance of the learning model.
Ahmad Faraz Khan et al. \cite{Khan2023PIFLPA} proposed PI-FL, a one-time personalized solution accompanied by a token-based incentive mechanism to reward personalized training. PI-FL outperforms other state-of-the-art methods by generating high-quality personalized models while respecting client privacy.
Mengqian Li et al. \cite{Li2024IMFLAI} combined game theory with an incentive mechanism and differential privacy (DP) to design an FL scheme (incentive mechanism for FL). They explored three DP variants and conducted convergence analysis on DP-based FL models. Then, using game theory, they analyzed the natural state of servers and clients during the FL process and formulated utility functions for both parties, considering potential attacks.
Jiawen Kang et al. \cite{Kang2023BlockchainEmpoweredFL} designed a user-centric privacy-preserving framework for decentralized federated learning (FL) in the healthcare metaverse. They used Age of Information (AoI) as an effective data freshness indicator and proposed an AoI-based contract theory model under Prospect Theory (PT) to incentivize data sharing in a user-centric manner.
Shenglv Zhang et al. \cite{Zhang2022ReinforcementLB} proposed a learning-based federated meta-learning incentive mechanism to motivate local clients to join data alliances. First, they suggested rewarding clients based on the amount of data they contribute to model training. To analyze the behavior of model owners and local clients, they formulated the incentive training task as a Stackelberg game and designed a reinforcement learning (RL) approach to learn the optimal pricing and participation strategies for task publishers and local clients.
Shuyuan Zheng et al. \cite{Zheng2021IncentiveMF} addressed both incentive mechanisms and privacy protection with FL-Market. This market incentivizes data owners by providing appropriate payments and privacy protection. FL-Market enables data owners to receive compensation based on their privacy loss.
Yufeng Zhan et al. \cite{Zhan2020ALI} studied incentive mechanisms in federated learning to encourage edge clients to participate in model training. Specifically, they designed a deep reinforcement learning (DRL)-based incentive mechanism to determine the optimal pricing strategy for the parameter server and the optimal training strategy for edge clients.
Jingwen Zhang et al. \cite{Zhang2021IncentiveMF} proposed a federated learning incentive mechanism based on reputation and reverse auction theory. Participants bid for tasks, with their reputation indirectly reflecting their reliability and data quality. In this federated learning scheme, participants are selected and rewarded by combining their reputation and bids within a limited budget.
Rui Hu et al. \cite{Hu2020TradingDF} designed an effective incentive mechanism using game theory to select users most likely to provide reliable data and compensate them for their privacy leakage costs.
Yajing Xu et al. \cite{Xu2023BESIFLBS} proposed a blockchain-enabled secure and incentive federated learning (BESIFL) paradigm. Specifically, BESIFL leverages blockchain to create a fully decentralized federated learning system, integrating effective mechanisms for malicious client detection and incentive management within a unified framework.
There are currently many incentive mechanisms for local training clients in federated learning. However, there are fewer incentive mechanisms specifically for personalized federated learning, and mechanisms that address the unique characteristics of personalized federated learning are particularly rare.
\section{PRELIMINARY}
\label{PRELIMINARY}
\subsection{Federated Learning}
\label{Federated Learning}



Federated learning was introduced to address privacy and security concerns associated with training data during the model training process. In federated learning, the training data remains stored on local clients, and a global model is trained by sharing model parameters. This approach ensures the privacy and security of the training data.

During the model training phase, different training clients may have varying amounts of data for different categories. Specifically, for training client $\mathbb{T}_\epsilon, \epsilon \in \{1, \ldots, M\}$, the data distribution may differ from other clients, $\mathbb{D}_{\mathbb{T}}^{j}$ $\neq$ $\mathbb{D}_{\mathbb{T}}^{k}$,$j,k \in \{1, \ldots, N\} \cap  (j\neq k) $. All training clients, however, use the same model architecture to train a global model $\mathfrak{M}(\omega;x)$. In the classic federated learning aggregation algorithm, FedAvg, the objective function of the global model is:
$$
\mathop{argmin}\limits_{\omega \in \mathbb{R}^d} \sum\limits_{i=1}^{N} \frac{|\mathbb{D}_i|}{n}\mathfrak{F}_{s}(\mathfrak{M}(\omega;x),y)
$$
where $\omega$ represents the model parameters of $\mathfrak{M}$, $N$ denotes the number of participating clients, and $n$ signifies the total number of instances across all training clients. $\mathfrak{F}_{s}$ stands for the optimization function of any supervised learning task. Due to potential data heterogeneity among different clients, training a single model by sharing model parameters in federated learning may no longer meet the needs of all training clients.
\subsection{Prototype}


When there are significant differences in data distribution among training clients in federated learning, training a single global model to learn the knowledge embedded in all clients' data is often ineffective. A deep learning model can be divided into two parts: the representation layer and the decision layer. The representation layer, typically comprising the early layers of a deep learning model, learns and expresses the knowledge contained in the training data for use by the decision layer. The decision layer classifies the input data based on the knowledge provided by the representation layer.

For training clients with non-IID data distributions, the representation layer learns different knowledge due to the varied data distributions and quantities of different types of data they possess. In this scenario, aggregating the representation layer parameters from different training clients into a single model, as done in traditional federated learning, fails to accurately fit the data from different clients.

A prototype is a high-level representation of the knowledge contained in the training client's data \cite{santos2013augmented}. For a class set $\mathfrak{C} = \{\mathfrak{c^1}, \mathfrak{c^2},\ldots,  \mathfrak{c^\mathbb{K}}\}$,, a common approach to obtain the prototypes for different categories is to compute the average of the embedding vectors of data points belonging to the same category. The prototype of $\mathfrak{c^\mathbb{K}}$ can be represented as:

$$
\mathfrak{V^\mathbb{K}} = \frac{1}{\mathbb{Y}_\mathbb{K}} \sum\limits_{x_{j}\in \mathbb{Y}_\mathbb{K}} \mathbb{F}_{\mathbb{em}}(x_{j})
$$
where $\mathfrak{V^\mathbb{K}} \in \mathbb{R}^D$ represents the prototype vector of dimension $D$ for category $\mathbb{K}$, $\mathbb{Y}_\mathbb{K}$ denotes the subset of data belonging to category $\mathbb{K}$, and $x_j \in \mathbb{R}^Z$ represents the input data of dimension $Z$. Therefore, $\mathbb{F}_{\mathbb{em}(x_{j})}:\mathbb{R}^Z \rightarrow \mathbb{R}^D$. 

\section{METHODS}
\label{METHODS}

In this section, we introduce the proposed BFLN model, which comprises two main components: the prototype-based aggregation algorithm (PAA) and the consensus algorithm based on cluster centroids (CACC). In the PAA, we propose a new federated learning aggregation algorithm specifically designed for non-IID data. In the CACC, we introduce a novel consensus algorithm tailored for federated learning, along with an incentive mechanism for personalized federated learning.

\subsection{Model Overview}
\label{Model Overview}
BFLN's design has the following goals:
\begin{enumerate}
	\item Improve the performance of federated learning models in non-IID scenarios.
	\item Provide a sustainable and reliable incentive mechanism for personalized federated learning.
\end{enumerate}

\begin{figure*}[htbp]
	\centering
	\includegraphics[width=.7\textwidth]{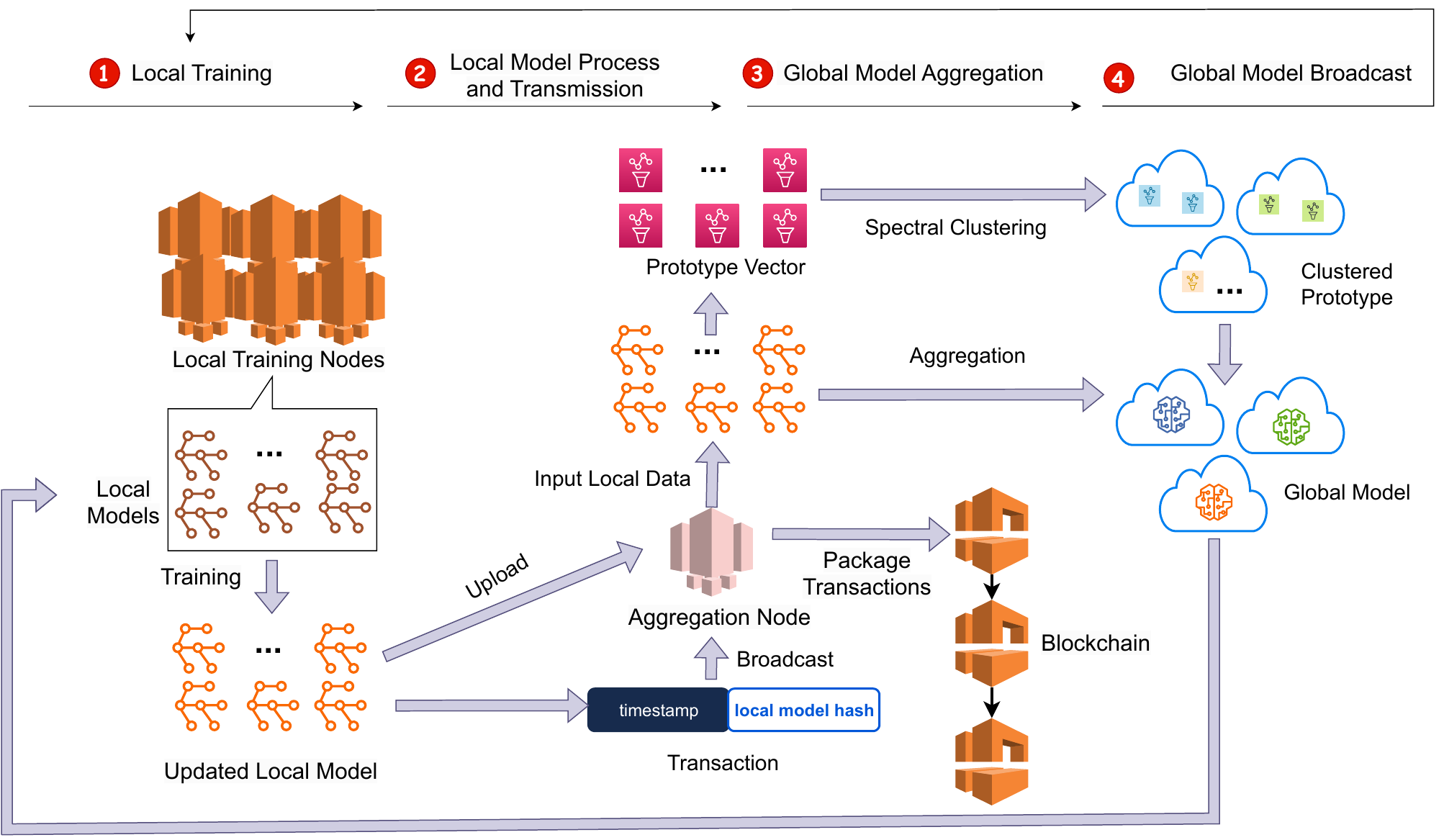}
	\caption{BFLN Model Structure}
	\label{BFLN Model Structure}
\end{figure*}

Figure \ref{BFLN Model Structure} illustrates the steps of the BFLN process:

\begin{enumerate}
	\item Local training clients train their local models.
	\item Local training clients send their local models to the aggregation client and submit the hash value of the local model to the blockchain.
	\item The aggregation client inputs sampled local data into the local models to obtain prototype vectors from different local models for the same data.
	\item The aggregation client spectral clusters the prototype vectors based on the aggregation algorithm and uses the FedAvg algorithm to aggregate local models with prototype vectors in the same cluster.
	\item The aggregation client submits the hash values of the local models involved in the aggregation to the blockchain and returns the aggregated model parameters to the local training clients.
	\item The consensus algorithm compares the hash values of the local models submitted by the local training clients with those submitted by the aggregation client, and rewards the local training clients whose submitted hash values match.
	
\end{enumerate}

\subsection{Prototype-based Aggregation Algorithm(PAA)} 

In the aggregation algorithm, the aggregator client first selects $\psi$ data belonging to the same category from local data. These data  are then fed into different local models, and the average of the resulting representation vectors for the same input data is computed to obtain a sampled prototype vector for each local model. 
Compared to \cite{Zhang2023AEP}, our method inputs the same data into different local models from the aggregation client to obtain prototype vectors. Additionally, we use Pearson correlation coefficient instead of cosine similarity to reflect the strength of similarity between prototype vectors.
For the $\mathfrak{e}$-th round of local training client $\mathbb{LT}^{\chi}$, its sampled representation vector is:
\begin{equation}
	\mathfrak{V}_{\chi}^{\mathfrak{e}}( x) = \frac{\sum\limits_{i=1}^{\psi}\mathbb{LM}_{\chi}^{\mathfrak{e}}(x_i)}{\psi}
\end{equation}
Hence, $\mathbb{LM}_{\chi}^{\mathfrak{e}}(x_i)$ represents the representation vector obtained by local model $\mathbb{LM}_{\chi}^{\mathfrak{e}}$ when input $x_i$ is provided. $\mathfrak{V}_{\chi}^{\mathfrak{e}}( x)$ denotes the prototype vector obtained after input $x$ is processed by the aggregation client.

Using the Pearson correlation coefficient, we calculate the linear similarity between different prototype vectors belonging to the same category. Suppose we have current local models $\mathbb{LM}_{\chi}^{\mathfrak{e}}$ and $\mathbb{LM}_{\delta}^{\mathfrak{e}}$, with their corresponding prototype vectors $\mathfrak{V}_{\chi}^{\mathfrak{e}}( x) $ and  $\mathfrak{V}_{\delta}^{\mathfrak{e}}( x) $, the Pearson correlation coefficient between $\mathfrak{V}_{\chi}^{\mathfrak{e}}( x) $ and  $\mathfrak{V}_{\delta}^{\mathfrak{e}}( x) $ is calculated as:

\begin{equation}
	\begin{aligned}
		&\mathfrak{S}_{\mathfrak{V}_{\chi}^{\mathfrak{e}}( x)  |\mathfrak{V}_{\delta}^{\mathfrak{e}}( x) } \\
		&= \frac{cov( \mathfrak{V}_{\chi}^{\mathfrak{e}}( x) ,\mathfrak{V}_{\delta}^{\mathfrak{e}}( x) )}{\sigma_{\mathfrak{V}_{\chi}^{\mathfrak{e}}( x) } \sigma_{\mathfrak{V}_{\delta}^{\mathfrak{e}}( x) }} \\
		& =  \frac{E[(\mathfrak{V}_{\chi}^{\mathfrak{e}}( x) - \mu_{\mathfrak{V}_{\chi}^{\mathfrak{e}}( x) })(\mathfrak{V}_{\delta}^{\mathfrak{e}}( x) -\mu_{\mathfrak{V}_{\delta}^{\mathfrak{e}}( x) })]}{\sigma_{\mathfrak{V}_{\chi}^{\mathfrak{e}}( x) } \sigma_{\mathfrak{V}_{\delta}^{\mathfrak{e}}( x) }}
	\end{aligned}
\end{equation}


Where $cov(\mathfrak{V}_{\chi}^{\mathfrak{e}}(x) ,\mathfrak{V}_{\delta}^{\mathfrak{e}}(x))$ denotes the covariance between $\mathfrak{V}_{\chi}^{\mathfrak{e}}(x)$ and $\mathfrak{V}_{\delta}^{\mathfrak{e}}(x)$, $\sigma_{\mathfrak{V}_{\chi}^{\mathfrak{e}}(x)}$ and  $\sigma_{\mathfrak{V}_{\delta}^{\mathfrak{e}}(x)}$ represent the standard deviations of $\mathfrak{V}_{\chi}^{\mathfrak{e}}(x)$ and $\mathfrak{V}_{\delta}^{\mathfrak{e}}(x) $ respectively. Using the Pearson correlation coefficient between   $\mathfrak{V}_{\chi}^{\mathfrak{e}}(x) $ and $\mathfrak{V}_{\delta}^{\mathfrak{e}}(x) $ measures the similarity between $\mathbb{LM}_{\chi}^{\mathfrak{e}}$ and  $\mathbb{LM}_{\delta}^{\mathfrak{e}}$ because we consider that if $\mathbb{LM}_{\chi}^{\mathfrak{e}}$ and $\mathbb{LM}_{\delta}^{\mathfrak{e}}$ are similar, their prototype vectors exhibit linear similarity. Unlike cosine similarity, the Pearson correlation coefficient can reflect the strength of similarity between prototype vectors, not just their direction. Using the above formula, we can obtain the Pearson correlation coefficient matrix between the prototype vectors corresponding to the sampled data at aggregation client.

\begin{equation}
	\Xi^{\mathfrak{e}} =  \left[ \begin{matrix}
		
		1 & \mathfrak{S}_{(1,2)}^{\mathfrak{e}} & \cdots & \mathfrak{S}_{(1,m)}^{\mathfrak{e}} \\
		
		\mathfrak{S}_{(2,1)}^{\mathfrak{e}} & 1 & \cdots & \mathfrak{S}_{(2,m)}^{\mathfrak{e}} \\
		
		\cdots &  \cdots & \cdots & \cdots \\
		
		\mathfrak{S}_{(m,1)}^{\mathfrak{e}} & \mathfrak{S}_{(m,2)}^{\mathfrak{e}} & \cdots & 1 \\
		
	\end{matrix} \right]
\end{equation}


According to the Pearson correlation coefficient matrix, we perform spectral clustering on the local models. Spectral clustering divides the local models into $\mathfrak{C}$ specified clusters based on $\Xi_\mathfrak{V^\mathbb{K}}$ , where each cluster contains an indeterminate number of local models. The prototype vectors obtained by models within each cluster exhibit stronger linear correlations. Using the FedAvg aggregation algorithm, we compute the average of local models within the same cluster to obtain multiple distinct global models. These global models are then returned to the local training clients within the cluster for the next round of model training.


\subsection{Consensus Algorithm based on Cluster Centroids (CACC)}


In BFLN, we propose a consensus algorithm based on clustering centroids. Leveraging the partition results from spectral clustering in PAA, we select local training clients closest to the centroids of each cluster and add them to the packing queue. Inspired by the DPoS consensus algorithm, where cluster centroids are added to the packing queue as block producers who take turns to pack blocks, in each round, these packing clients also serve as aggregation clients by inputting local data into different local models to obtain prototype vectors. This queuing mechanism ensures block packing efficiency, reducing computational waste compared to the widely used PoW consensus algorithm. Given their randomness and representative nature of cluster clients, centroids are selected from each cluster to join the packing queue. 
After performing spectral clustering on local models in PAA, we compute the centroids of each cluster. We calculate the average Pearson correlation coefficient within each cluster to determine the centroids. For the $\tau$th cluster $\mathfrak{C}_{\tau}^{\mathfrak{e}}$ in the $\mathfrak{e}$-th round, its centroid is given by:
\begin{equation}
	\mathfrak{c}_{\tau}^{\mathfrak{e}} = \frac{\sum\limits_{\upsilon = 1}^{\mathfrak{N}_{\tau}^{\mathfrak{e}}} \mathfrak{S}_{\upsilon}^{\mathfrak{e}}}{\mathfrak{N}_{\tau}^{\mathfrak{e}}}
\end{equation}

Calculate the Euclidean distance between each point in the cluster and  $\mathfrak{c}_{\tau}^{\mathfrak{e}}$. For the $\theta$-th point  $\mathfrak{p}_{\tau;\theta}^{\mathfrak{e}}$ in cluster  $\mathfrak{C}_{\tau}^{\mathfrak{e}}$, the Euclidean distance $\mathfrak{L}_{\theta, \mathfrak{c}}$ to $\mathfrak{c}_{\tau}^{\mathfrak{e}}$ is:

\begin{equation}
	\mathfrak{D}_{ \theta, \mathfrak{c}} = \mathfrak{p}_{\tau;\theta}^{\mathfrak{e}} - \mathfrak{c}_{\tau}^{\mathfrak{e}}
\end{equation}

\begin{equation}
	\mathfrak{L}_{\theta, \mathfrak{c}} = \sqrt{\sum_{i=1}^{\mathfrak{N}_{\mathfrak{D}_{ \theta, \mathfrak{c}}}} \mathfrak{d}_{i}^2}
\end{equation}

Where $\mathfrak{N}_\theta$  denotes the number of elements in $\mathfrak{D}_{ \theta, \mathfrak{c}}$, and $\mathfrak{d}_{i}$  represents an element in $\mathfrak{D}_{ \theta, \mathfrak{c}}$.


We select the point in cluster $\mathfrak{C}_{\tau}^{\mathfrak{e}}$ with the smallest $\mathfrak{L}_{\mathfrak{c}, \theta}$, corresponding to its local training client as the centroid of the cluster, which enters the packaging queue.

\subsubsection{Incentive mechanism based on cluster membership size}

Assuming each group has $n_i$ members, with $j$ groups in total and a total of $N$ members, where $\sum\limits_{i=1}^j n_i = N$, and the total reward is $\mathfrak{R}$. We have designed an allocation function $\Gamma(n_i)$ that increases with the group size $n_i$, with the rate of increase accelerating, ensuring that the per capita reward increases with the size of the group. Different groups receive varying amounts of rewards, while individuals within the same group receive equal shares of the reward.

\begin{equation}
	\Gamma(n_i) = \kappa {n_i}^{\rho}, \rho>1
	\label{Gamma}
\end{equation}

\begin{equation}
	\kappa = \frac{\mathfrak{R}} {\sum\limits_{i=1}^j n_i^\rho}
\end{equation}

The reward allocated to training client $\mathfrak{T_k}$ is given by $\mathfrak{r_k} =\frac{\Gamma(n_i)}{n_i}$.
When a training client $\mathfrak{T_k}$ sends its local training model to the aggregation client and requests aggregation, the training client incurs a cost $\mathfrak{g_k}$ as compensation.  $\mathfrak{g_k}$ will be reward to aggregation node.
\begin{equation}
	\mathfrak{g_k} = \frac{\kappa}{N}
\end{equation}

When a client initially joins the blockchain network, it receives a certain amount of tokens from the blockchain network to pay for aggregation client fees. 

To prevent scenarios where local training clients attempt to claim rewards without participating in training, after sending their local models to the aggregation client, local training clients submit a transaction to the blockchain network containing the hash value of their local model. Upon receiving local models, during block packaging, the aggregation client includes transactions in the block that contain hash values of all aggregated local models. The consensus algorithm verifies whether the hash values of local models included in the transactions initiated by the aggregation client match those submitted by the local training clients. Only if the aggregation client's transactions include the hash values of local models can the local training clients receive rewards distributed by the blockchain network.
\section{Experiments}
\label{Experiments}
\subsection{Runtime Environment}
Our experiments were conducted on a server, using PyTorch version 2.3.0, CUDA version 12.4, with a CentOS
equipped with 395GB of RAM and a 24TB hard disk, the GPUs in the server are  Quadro RTX 5000 (with a VRAM of 16GB). The parameters are shown in table \ref{Model Parameters}.

\begin{table}[htbp]
	\centering
	\caption{Model Parameters}
	\begin{tabularx}{\columnwidth}{Xr}
		\toprule
		parameter & value \\
		\midrule
		Batch size of client's local training& 64  \\
		Learning rate  of client's local training& 0.001  \\
		Local epochs  of client's local training& 5  \\
		Max running round  of client's local training& 50 \\
		Number of aggregation clients  (in BFLN) & 1 \\
		Number of local training clients & 20 \\
		Initial stake for blockchain clients& 5 \\
		Local training total stake reward& 20 \\
		$\rho$ in Equation \ref{Gamma}& 2 \\
		\bottomrule
	\end{tabularx}
	\label{Model Parameters}
\end{table}

\subsection{Dataset}
\begin{itemize}
	\item \textbf{CIFAR10}: This dataset contains RGB color images classified into 10 categories: airplane, automobile, bird, cat, deer, dog, frog, horse, ship, and truck. Each image is 32 × 32 pixels. There are 6,000 images per category, totaling 50,000 training images and 10,000 testing images.
	\item \textbf{CIFAR100}: This dataset consists of RGB color images divided into 100 categories, with each category containing 600 images sized 32 × 32 pixels. Of these, 500 images are used for training and 100 for testing. Each image in CIFAR100 has two labels: $fine\_labels$ for fine-grained classification and $coarse\_labels$ for coarse-grained classification.
	\item \textbf{SVHN}: The Street View House Numbers (SVHN) dataset is a real-world dataset with 10 classes. It includes 73,257 digits for training and 26,032 digits for testing. The dataset is available in two formats: original images with character-level bounding boxes and MNIST-like 32 × 32 images centered around a single character.
\end{itemize}

\subsection{Baseline}
\begin{itemize}
	
	\item FedAvg: proposed by McMahan  et al. in 2017 \cite{mcmahan2017communication}, is a vanilla aggregation algorithm in federated learning. It creates a global model by averaging the parameters of local models.
	\item FedHKD: introduced by Huancheng Chen et al. in 2023 \cite{chen2023best}, uses knowledge distillation to train local models. Each client extracts the mean representation and corresponding soft predictions of local data and sends this information to the server. Unlike other personalized federated learning methods based on knowledge distillation, FedHKD does not rely on public datasets or deploy generative models on the server. The server aggregates this hyper-knowledge information and broadcasts it back to clients to support their local training.
	\item FedProto: proposed by Yue Tan et al. in 2022 \cite{tan2022fedproto}, addresses heterogeneous data distribution across clients by aligning global prototypes with local prototypes. Additionally, FedProto resolves the issue of heterogeneous client model architectures by using the trained prototypes of the same category to perform inference tasks.
	\item FedProx:  introduced by Tian Li et al. in 2018 \cite{li2020federated}, can handle heterogeneous federated data while maintaining similar privacy and computational advantages. The proposed framework demonstrates better robustness and stability in heterogeneous federated networks.
\end{itemize}

The implementation of the baselines is based on https://github.com/lunan0320/Federated-Learning-Knowledge-Distillation.

\subsection{Accuracy}

\begin{table*}[htbp]
	\centering
	\resizebox{0.85\textwidth}{!}{
		\begin{tabular}{l|ccccccccc}
			Model & CIFAR10-0.1 & CIFAR10-0.3 & CIFAR10-0.5 & CIFAR100-0.1 & CIFAR100-0.3 & CIFAR100-0.5 & SVHN-0.1 & SVHN-0.3 & SVHN-0.5 \\ 
			\hline
			BFLN-Cluster-2 & 0.8635 & 0.7718 & 0.7399 & 0.5110 & 0.3628 & 0.3215 & 0.9540 & 0.9295 & 0.9208 \\\hline
			BFLN-Cluster-3 & 0.8626 & 0.7629 & 0.7491 & 0.5149 & 0.3716 & 0.3280 & 0.9619 & 0.9314 & 0.9199 \\\hline
			BFLN-Cluster-4 & 0.8584 & 0.7699 & 0.7536 & 0.5127 & 0.3683 & 0.3364 &  0.9551 & 0.9340 & 0.9202 \\\hline
			BFLN-Cluster-5 & \textbf{0.8769} & 0.7738 & 0.7458 & 0.5213 & 0.3875 & 0.3318 &  0.9592 & 0.9350 & 0.9204 \\\hline
			BFLN-Cluster-6 & 0.8707 & 0.7742 & 0.7542 & 0.5402 & 0.3936 & 0.3504 & 0.9594 &  \textbf{0.9351} & 0.9195 \\\hline
			BFLN-Cluster-7 & 0.8706 & \textbf{0.7806} & \textbf{0.7656} & \textbf{0.5489} &  \textbf{0.3973} &  \textbf{0.3589} &  \textbf{0.9620} & 0.9350 &  \textbf{0.9261} \\\hline
			FedAvg & 0.8579 & 0.7658 & 0.7438 & 0.5196 & 0.3703 & 0.3155 & 0.9576 & 0.9322 & 0.9205 \\\hline
			FedHKD & 0.1088 & 0.0982 & 0.7499 & 0.0101 & 0.0110 & 0.0099 & 0.0584 & 0.0605 & 0.0735 \\\hline
			FedProto & 0.8667 & 0.7620 & 0.7501 & 0.4393 & 0.3949 & 0.3336 & 0.9593 & 0.9336 & 0.9193 \\\hline
			FedProx & 0.8665 & 0.7738 & 0.7574 & 0.5297 & 0.3872 & 0.3465 & 0.9566 & 0.9310 & 0.9200 \\ \hline
			
	\end{tabular}
}
	\caption{ Model Accuracy in CIFAR10,CIFAR100 and SVHN with different local data distribution }
	
	\label{Accuracy Table}
\end{table*}

%
%
%
%

In this chapter, we tested the model performance on three datasets under three different levels of data label bias distribution. We compared the performance of our proposed BFLN with other federated learning models, specifically FedAvg, FedHKD, FedProto, and FedProx.

Table \ref{Accuracy Table} presents the accuracy of the proposed BFLN model with different numbers of clusters. In the CIFAR10 dataset, when the training data bias is 0.1, BFLN achieves the best performance with 5 clusters, reaching an accuracy of 0.8769, which is a 1.02\% improvement over the best-performing baseline, FedProto. When the data bias is 0.3, BFLN performs best with 7 clusters, achieving an accuracy of 0.7806, surpassing other baseline algorithms. At a bias level of 0.5, BFLN again achieves the best results with 7 clusters.

For the CIFAR100 dataset, when the training data bias is 0.1, BFLN with 7 clusters improves by 1.92\% over the best-performing baseline, FedProx, achieving the highest model accuracy. With a bias of 0.3, BFLN reaches an accuracy of 0.3973 with 7 clusters, higher than any other baseline algorithms. When the bias is 0.5, BFLN achieves an accuracy of 0.3589 with 7 clusters, surpassing the best-performing baseline, FedProx, by 1.24\%. In the CIFAR100 dataset, the best performance is consistently achieved with 7 clusters, which is the largest number of clusters we tested.

In the SVHN dataset, with a training data bias of 0.1, BFLN achieves the best results with 7 clusters. At a bias level of 0.3, the highest accuracy is achieved with 6 clusters. When the bias is 0.5, BFLN with 7 clusters outperforms all baseline models.

Compared to the CIFAR10 and CIFAR100 datasets, the accuracy improvement of the BFLN model on the SVHN dataset is less pronounced. This might be because the SVHN dataset is relatively simpler than CIFAR10 and CIFAR100, and the models already achieve relatively high accuracy under different bias levels. In CIFAR10 and CIFAR100, when the data bias is lower, the improvement of our proposed BFLN model over baseline models might be more significant. This could be because BFLN clusters the prototype vectors and aggregates local model parameters, resulting in better clustering of training clients with similar data when the data bias is higher.

\subsection{Reward Trends }

\begin{figure*}[htbp]
	\centering
	
	\begin{subfigure}[b]{0.32\textwidth}
		\centering
		\includegraphics[width=\textwidth]{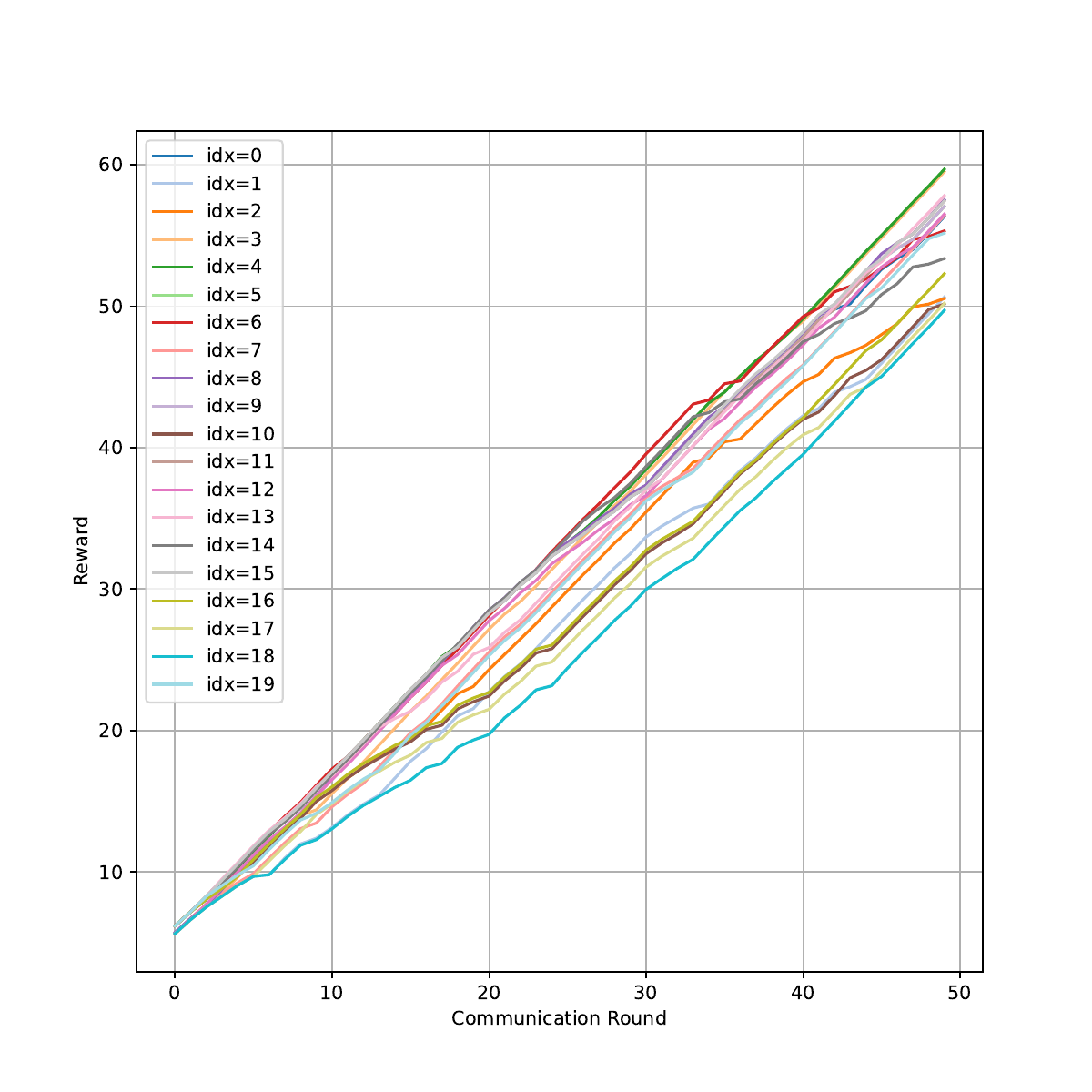}
		\caption{CIFAR10-2-0.1-reward}
		\label{CIFAR10-2-0.1-reward}
	\end{subfigure}
	\begin{subfigure}[b]{0.32\textwidth}
		\centering
		\includegraphics[width=\textwidth]{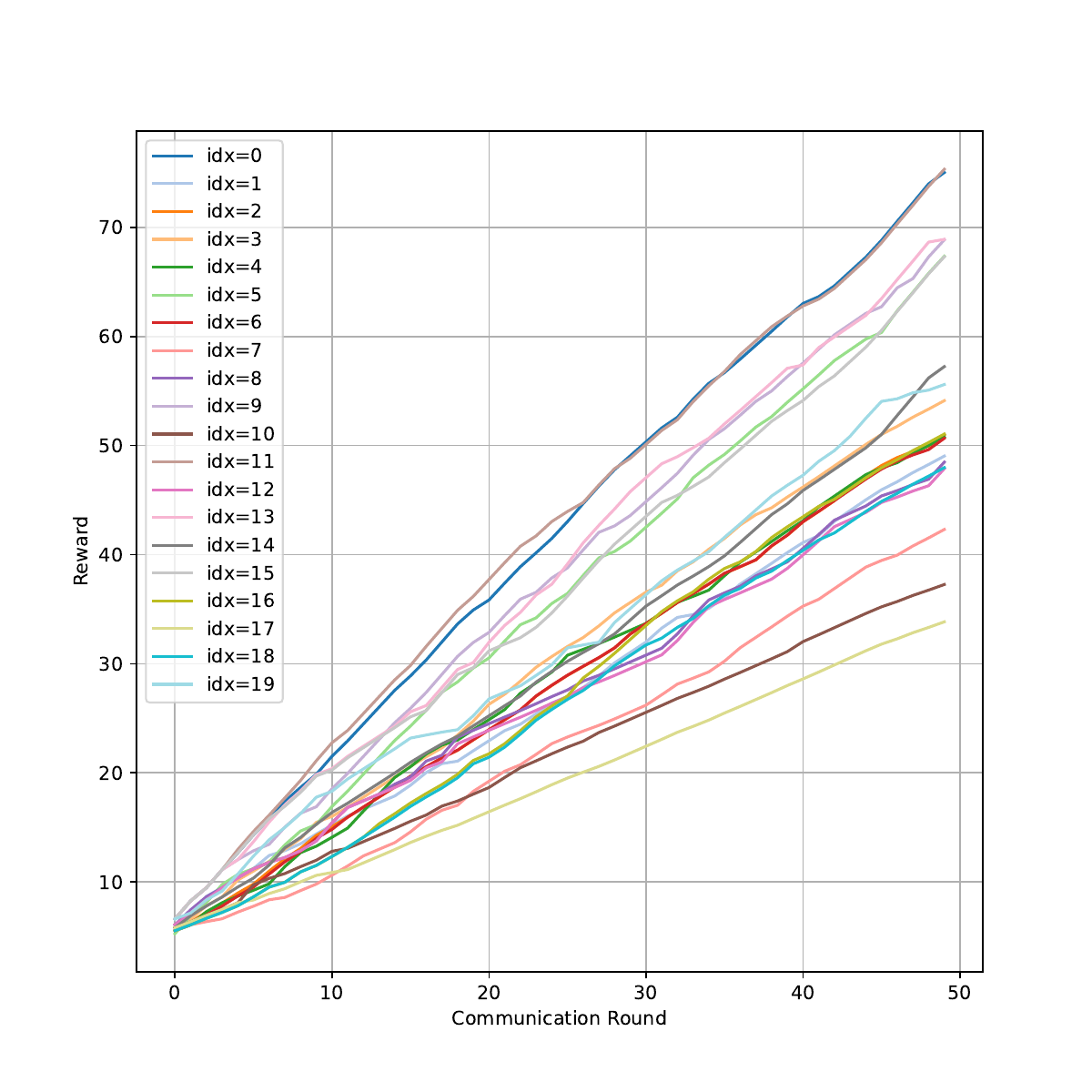}
		\caption{CIFAR10-7-0.1-reward}
		\label{CIFAR10-7-0.1-reward}
	\end{subfigure}
	\begin{subfigure}[b]{0.32\textwidth}
		\centering
		\includegraphics[width=\textwidth]{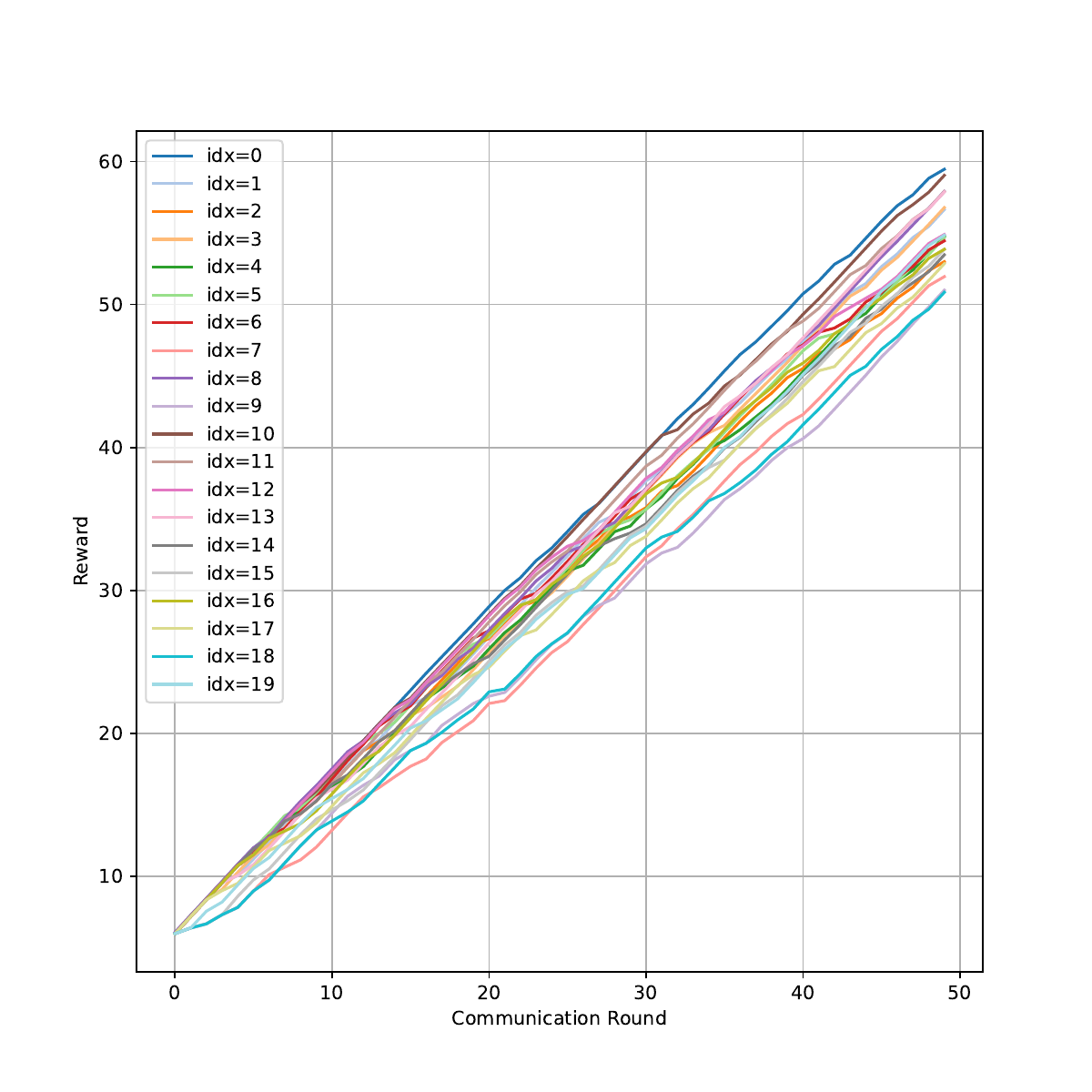}
		\caption{CIFAR100-2-0.1-reward}
		\label{CIFAR100-2-0.1-reward}
	\end{subfigure}
	
	\begin{subfigure}[b]{0.32\textwidth}
		\centering
		\includegraphics[width=\textwidth]{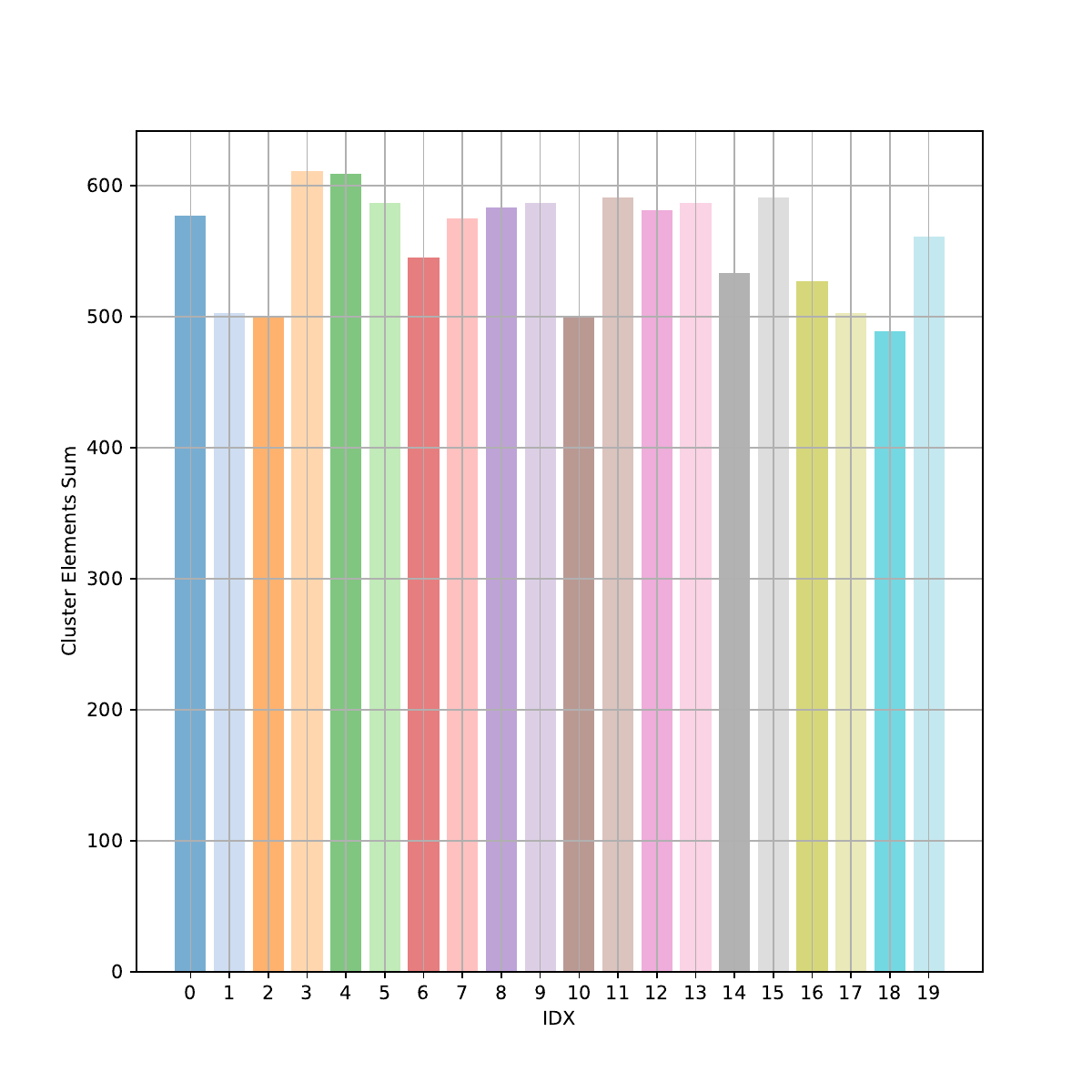}
		\caption{CIFAR10-2-0.1-cluster}
		\label{CIFAR10-2-0.1-cluster}
	\end{subfigure}
	\begin{subfigure}[b]{0.32\textwidth}
		\centering
		\includegraphics[width=\textwidth]{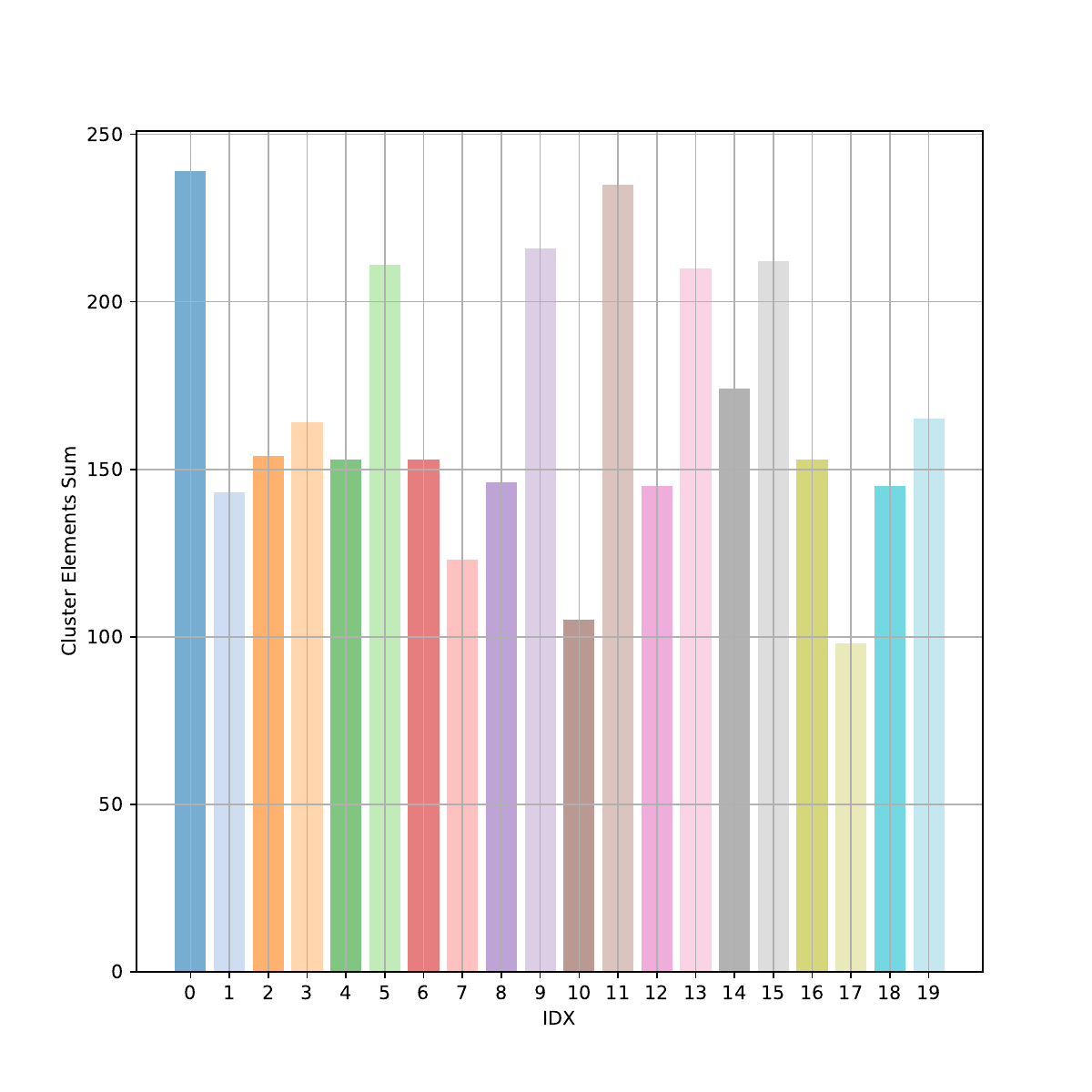}
		\caption{CIFAR10-7-0.1-cluster}
		\label{CIFAR10-7-0.1-cluster}
	\end{subfigure}
	\begin{subfigure}[b]{0.32\textwidth}
		\centering
		\includegraphics[width=\textwidth]{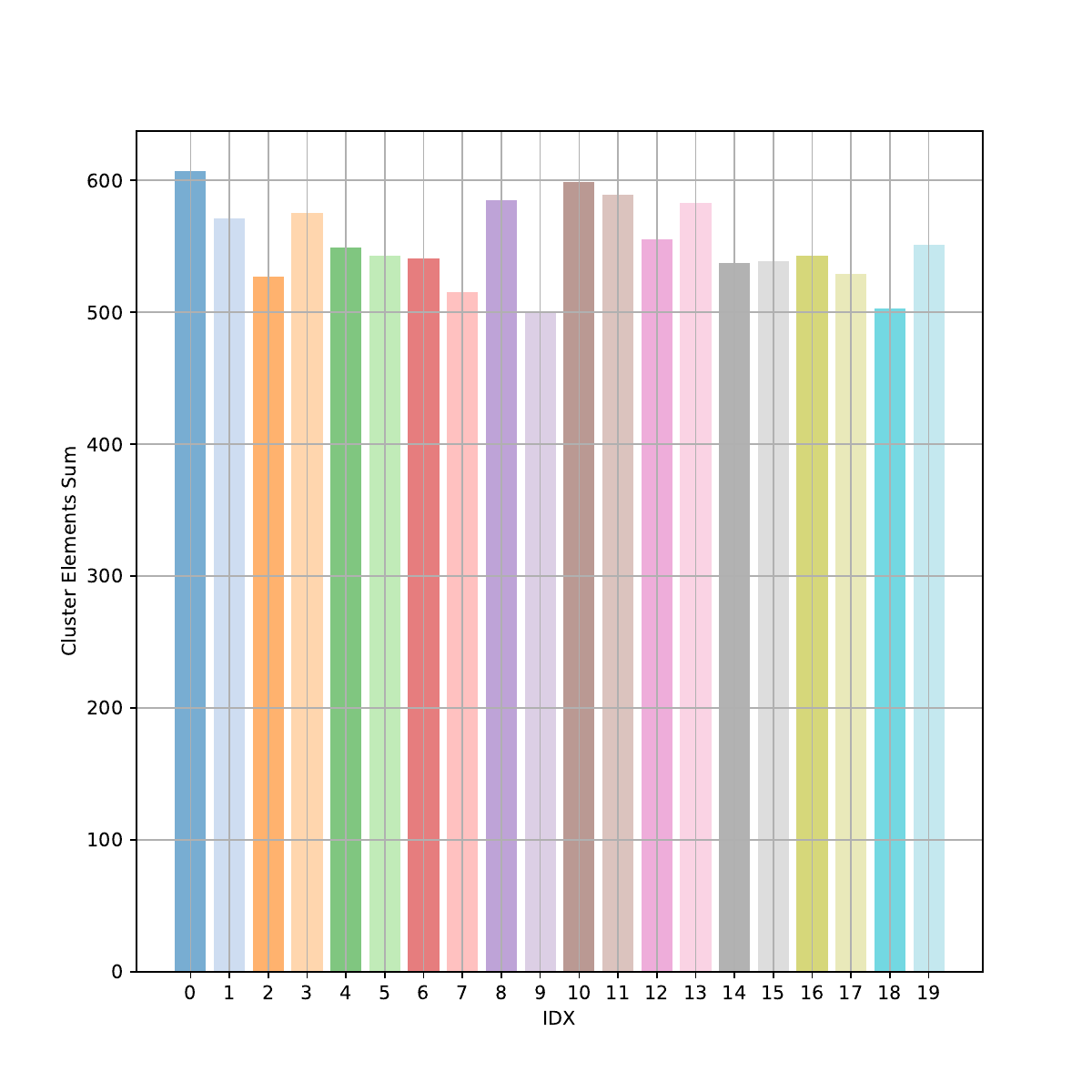}
		\caption{CIFAR100-2-0.1-cluster}
		\label{CIFAR100-2-0.1-cluster}
	\end{subfigure}
	
	\begin{subfigure}[b]{0.32\textwidth}
		\centering
		\includegraphics[width=\textwidth]{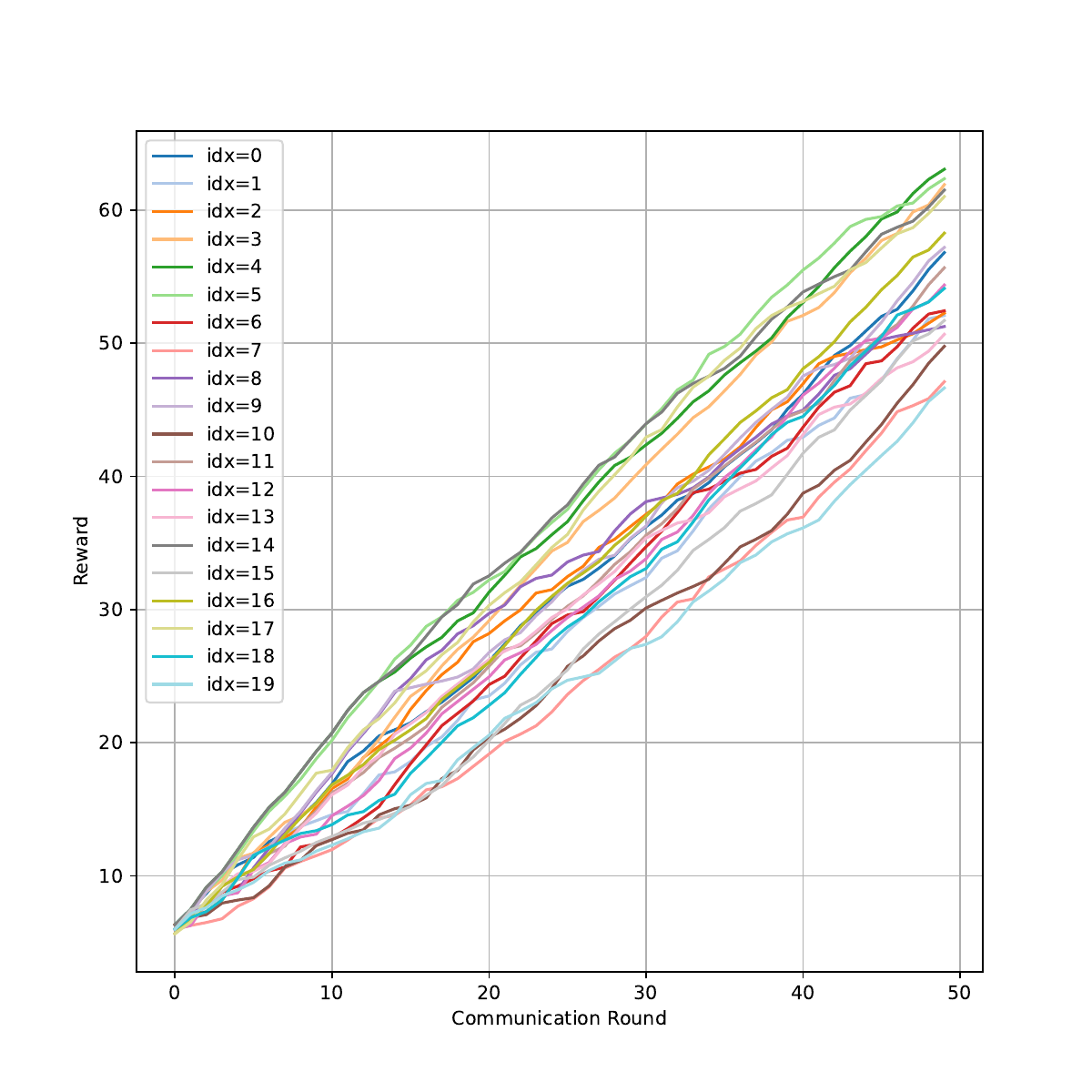}
		\caption{CIFAR100-7-0.1-reward}
		\label{CIFAR100-7-0.1-reward}
	\end{subfigure}
	\begin{subfigure}[b]{0.32\textwidth}
		\centering
		\includegraphics[width=\textwidth]{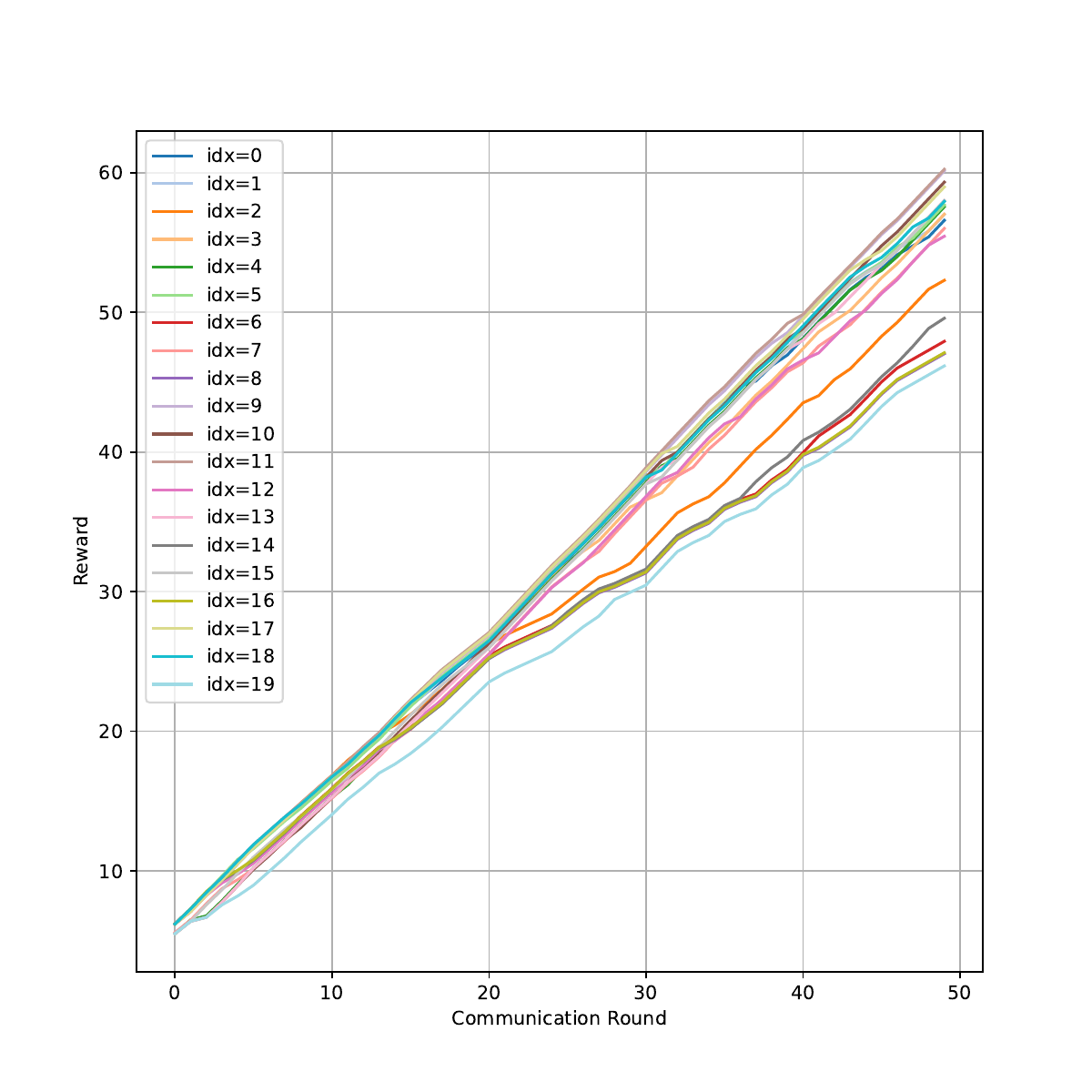}
		\caption{SVHN-2-0.1-reward}
		\label{SVHN-2-0.1-reward}
	\end{subfigure}
	\begin{subfigure}[b]{0.32\textwidth}
		\centering
		\includegraphics[width=\textwidth]{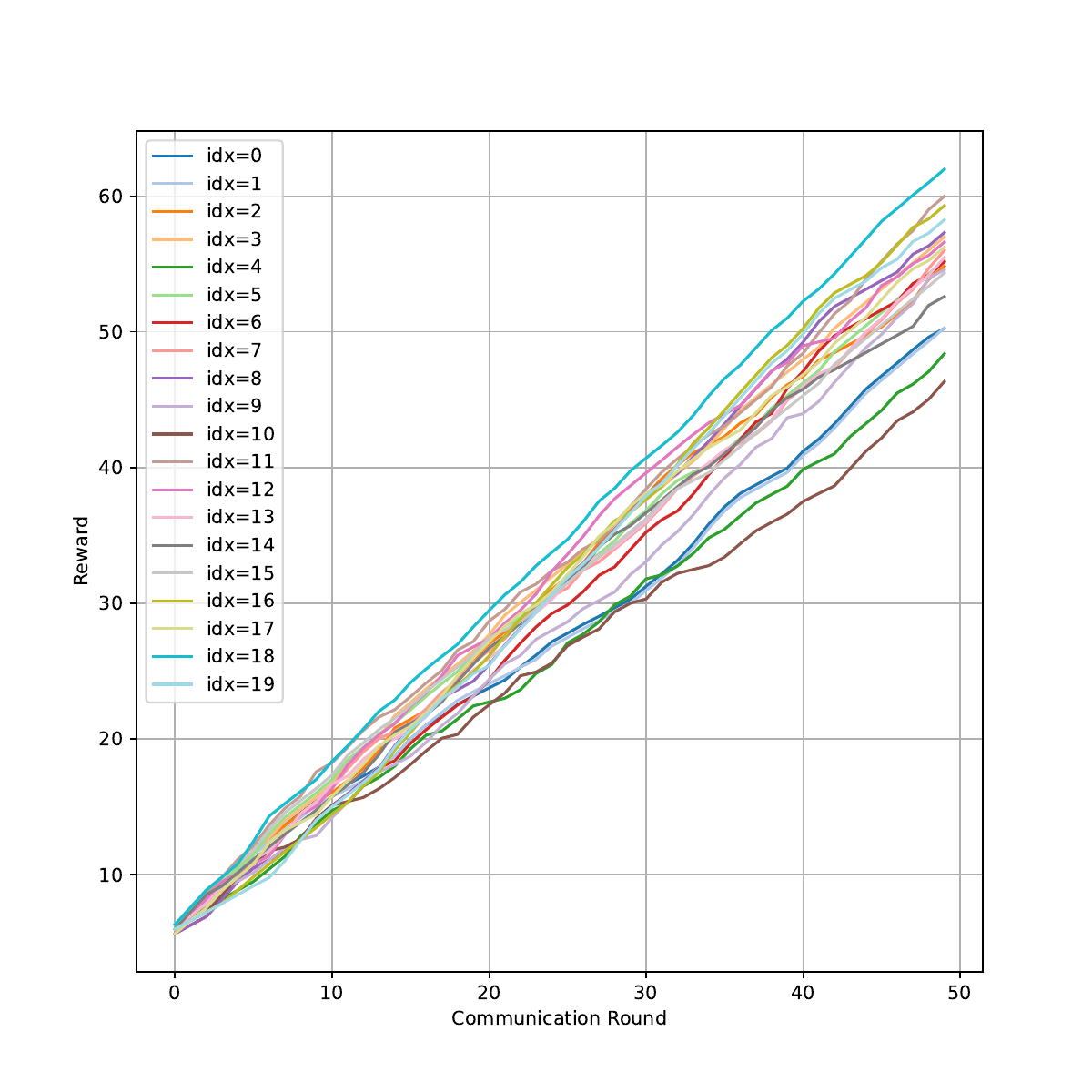}
		\caption{SVHN-7-0.1-reward}
		\label{SVHN-7-0.1-reward}
	\end{subfigure}

	\begin{subfigure}[b]{0.32\textwidth}
		\centering
		\includegraphics[width=\textwidth]{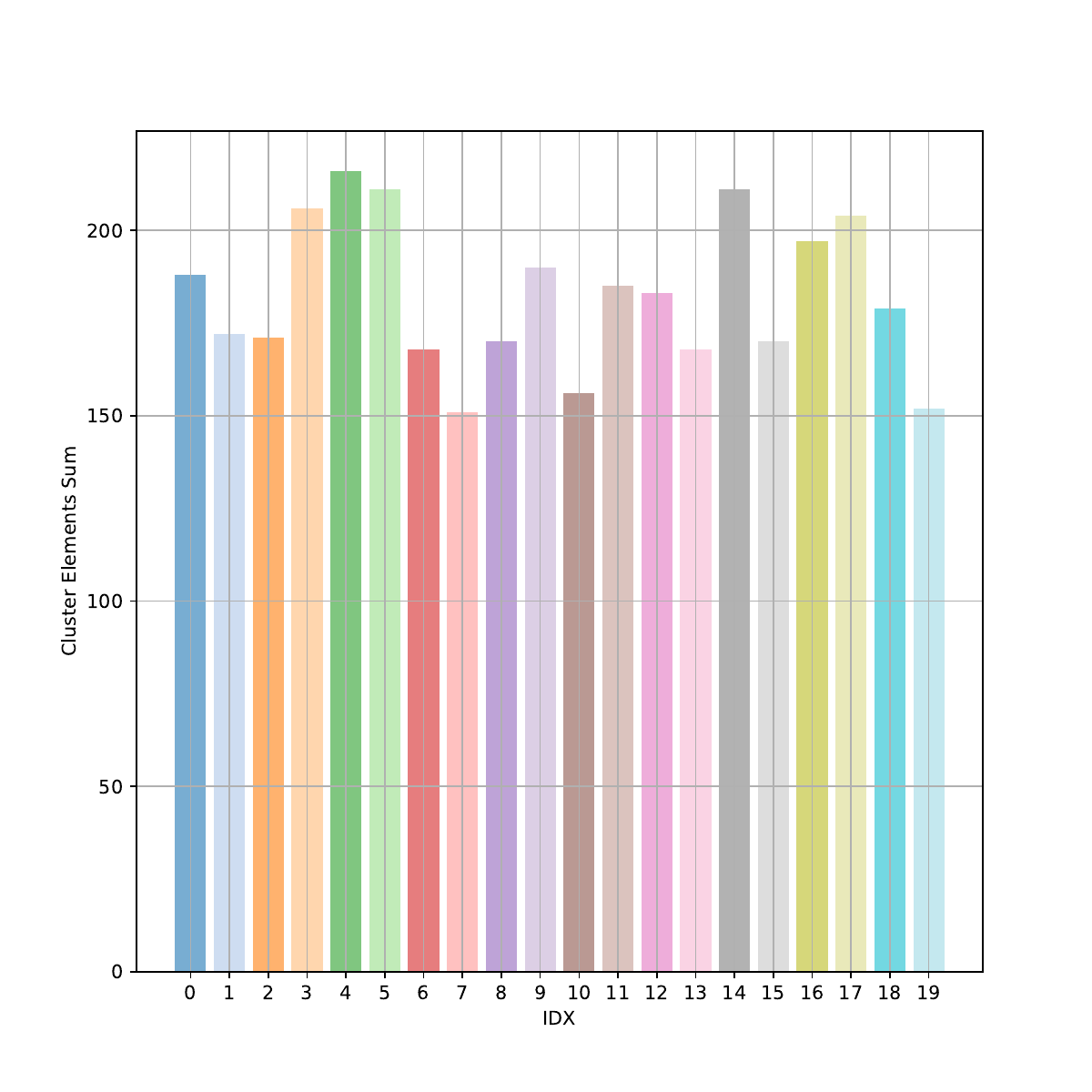}
		\caption{CIFAR100-7-0.1-cluster}
		\label{CIFAR100-7-0.1-cluster}
	\end{subfigure}
	\begin{subfigure}[b]{0.32\textwidth}
		\centering
		\includegraphics[width=\textwidth]{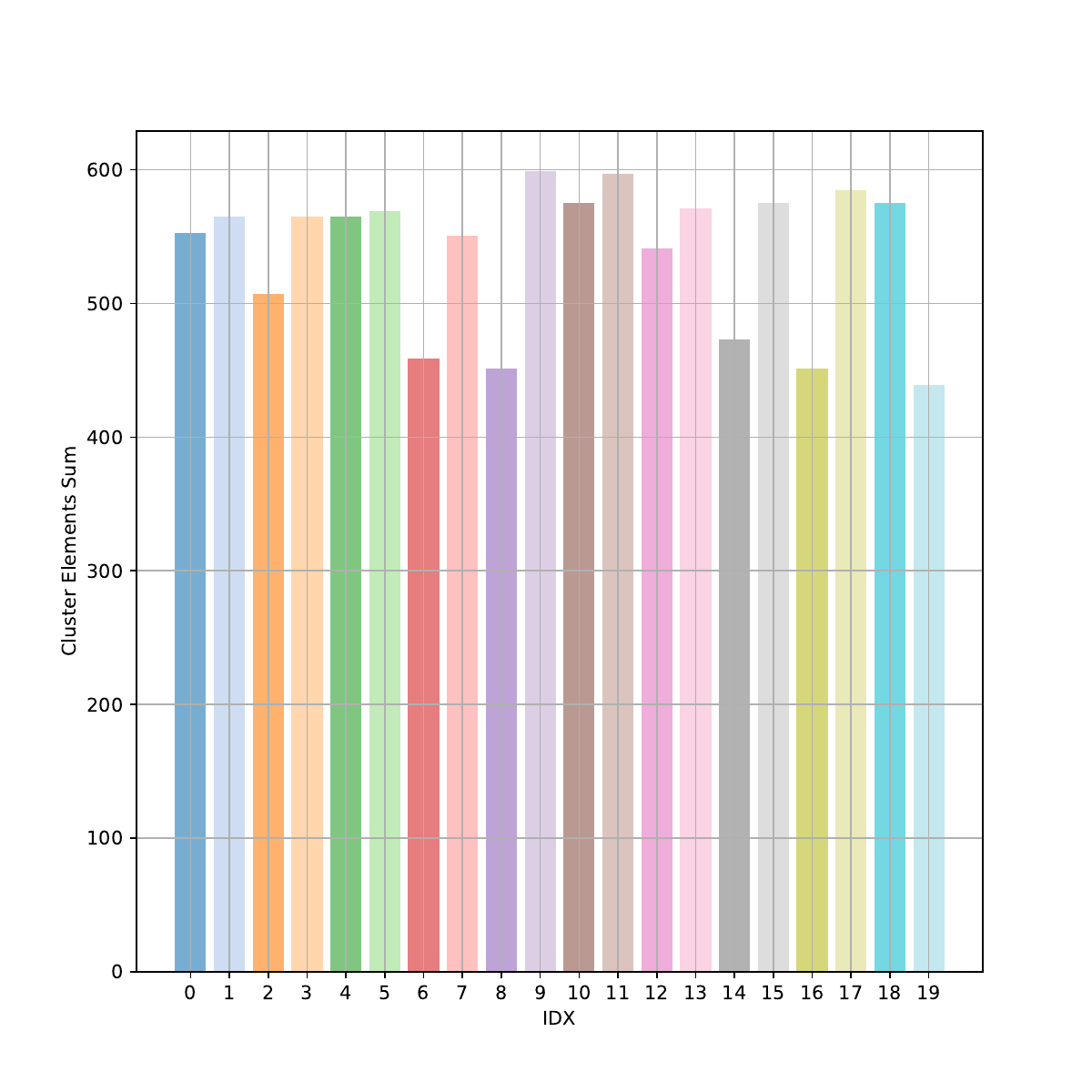}
		\caption{SVHN-2-0.1-cluster}
		\label{SVHN-2-0.1-cluster}
	\end{subfigure}
	\begin{subfigure}[b]{0.32\textwidth}
		\centering
		\includegraphics[width=\textwidth]{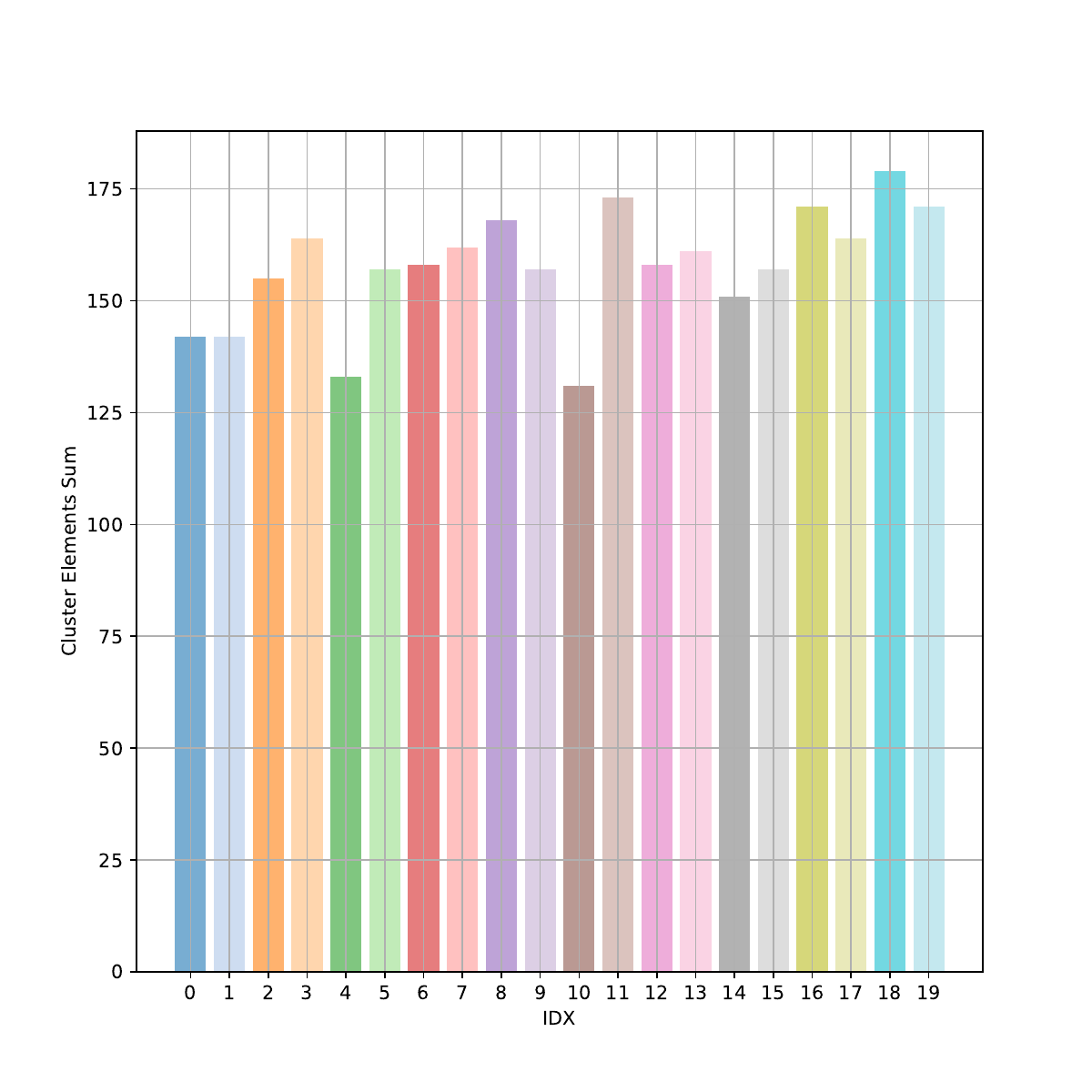}
		\caption{SVHN-7-0.1-cluster}
		\label{SVHN-7-0.1-cluster}
	\end{subfigure}

	\caption{BFLN Reward Trends and Different Local Training clients Cluster Summarize in Different Datasets }
	\label{BFLN Reward Trends}
\end{figure*}

%
%
%
%
%
%

In BFLN, we propose CACC, which includes an incentive mechanism for personalized federated learning. In CACC, we believe that during the clustering phase of aggregation, the more clients a local training client’s cluster contains, the greater its contribution to other local training clients. Therefore, the consensus algorithm will reward the clients in that cluster more to encourage their continued participation in the aggregation process.

In Figure 2, we show the rewards obtained by local training clients in our proposed BFLN model under different clustering numbers across three datasets. The subfigure titles connect the dataset name, clustering number, bias degree, and image type with hyphens. "Reward" indicates the reward trend graph, and "cluster" represents the sum of the clients in each cluster during the model training process.

In the CIFAR10 dataset, with 2 clusters and a bias degree of 0.1, as seen in Figure \ref{CIFAR10-2-0.1-reward}, the local training clients with indices 4 and 3 receive the most rewards by the end of model training. In Figure \ref{CIFAR10-2-0.1-cluster}, it can be observed that these clients also belong to clusters with a larger total number of clients. Conversely, clients with indices 1, 2, 10, 17, and 18 receive fewer rewards, and their corresponding cluster sizes are relatively smaller. With 7 clusters, Figure \ref{CIFAR10-7-0.1-reward} shows that clients with indices 0 and 11 receive the most rewards. Figure \ref{CIFAR10-7-0.1-cluster} indicates that the clusters containing clients with indices 2 and 11 have the highest total client count, while client 17 receives the fewest rewards and belongs to the smallest cluster.

For the CIFAR100 dataset, with 2 clusters, Figure \ref{CIFAR100-2-0.1-reward} shows that the local training client with index 0 receives the highest reward. Figure \ref{CIFAR100-2-0.1-cluster} corroborates this, showing the cluster containing client 0 has the largest total client count. clients with indices 18 and 9 receive the fewest rewards, and their cluster sizes are correspondingly smaller. With 7 clusters, Figure \ref{CIFAR100-7-0.1-reward} shows that client 4 receives the highest reward, and Figure \ref{CIFAR100-7-0.1-cluster} confirms that the cluster containing client 4 has the highest total client count. clients 7 and 19 receive the fewest rewards, with their cluster sizes being correspondingly smaller.

In the SVHN dataset, with 2 clusters, Figure \ref{SVHN-2-0.1-reward} shows that client 9 receives the most rewards, and Figure \ref{SVHN-2-0.1-cluster} confirms that this client’s cluster has the largest total client count. client 19 receives the fewest rewards, and its cluster is the smallest. With 7 clusters, Figure \ref{SVHN-7-0.1-reward} shows that client 18 receives the most rewards, and Figure \ref{SVHN-7-0.1-cluster} confirms that this client’s cluster has the largest total client count. client 10 receives the fewest rewards, and its cluster is the smallest.

Figure 2 shows that with fewer clusters, the distribution of rewards among clients is denser and less distinct. This is because, with fewer clusters, the similarity requirement for prototype vectors during clustering is lower, making it easier for local training clients to be grouped into the same cluster. With more clusters, the similarity requirement for prototype vectors is higher, and only clients with highly similar prototype vectors are grouped into the same cluster, leading to greater differences in rewards among different training clients.

\section{Conclusion}
\label{Conclusion and Future Work}
In this paper, we propose a Blockchain-based Federated Learning Model for Non-IID Data. Our goal is to address the non-IID problem and develop an incentive mechanism for personalized federated learning. To achieve this, we introduce two main components: the prototype-based aggregation algorithm and the consensus algorithm based on cluster centroids. These components cluster local models by comparing prototype vectors obtained from different local models with the same input data, thereby achieving personalized federated learning. Additionally, we propose an incentive mechanism tailored to the characteristics of BFLN, ensuring the enthusiasm of clients participating in federated learning training. Compared to existing personalized federated learning methods, our proposed BFLN demonstrates better performance in scenarios with non-IID local data distributions. Experimental results show that our incentive mechanism is effective and aggregation algorithm has better performance than baseline algorithms.

\section*{Acknowledgement}
This work was supported by the National Natural Science Foundation of China under Grant No. 62272024.

\end{sloppypar}
\end{document}